\documentclass[printer]{aa}
\usepackage{graphicx,natbib}
\usepackage{txfonts}
\usepackage{epic,eepic}

\begin{document}
 
%
%
\def\valid{}    

\font\caps=cmcsc10                  
\font\dunh=cmdunh10  at 12.0 true pt 
\font\dunhs=cmdunh10 
\font\vbold=cmbx10 scaled \magstep1 
\font\sevenbf=cmbx7
\font\sevenit=cmti7
\font\Kapi=cmr17

\def\MEV{DOME}
\def\RTE{equation of radiative transfer}
\def\etal{{et al}}
\def\HW{H\&W}
\def\OK{O\&K}
\def\ok{O\&K}
\def\RH{R\&H}

\def\ibmrs{\hbox{\tt RS/6000}}
\def\hp{\hbox{\tt HP~9000}}
\def\dec{\hbox{\tt DEC~5000}}
\def\axp{\hbox{\tt AXP}}
\def\ibmmf{\hbox{\tt IBM~3090}}
\def\ibmpc{\hbox{\tt 486DX}}
\def\cray{\hbox{\tt Cray 2}}
\def\ymp{\hbox{\tt YMP}}
\def\nec{\hbox{\tt NEC}}

\def\g{\gamma}
\def\b{\beta}
\def\m{\mu}
\def\e{\epsilon}
\def\n{\nu}
\def\l{\lambda}
\def\L{\Lambda}
\def\t{\tau}
\def\pder#1#2{{\partial #1 \over \partial #2}}
\def\div#1#2{{#1\over #2}}
\def\rout{\ifmmode{r_{\rm out}}\else\hbox{$r_{\rm out}$}\fi}
\def\tmax{\ifmmode{\tau_{\rm max}}\else\hbox{$\tau_{\rm max}$}\fi}
\def\tstd{\ifmmode{\tau_{\rm std}}\else\hbox{$\tau_{\rm std}$}\fi}
\def\vmax{\ifmmode{v_{\rm max}}\else\hbox{$v_{\rm max}$}\fi}
\def\muE{\ifmmode{\mu_{\rm E}}\else\hbox{$\mu_{\rm E}$}\fi} 
\def\pE{\ifmmode{p_{\rm E}}\else\hbox{$p_{\rm E}$}\fi} 
\def\bmax{\ifmmode{\b_{\rm max}}\else\hbox{$\b_{\rm max}$}\fi}
\def\kms{\hbox{$\,$km$\,$s$^{-1}$}}
\def\ergs{\hbox{$\,$erg$\,$s$^{-1}$}}
\def\kpc{\hbox{$\,$kpc} }
\def\ang{\hbox{\AA}}
\def\Msun{\hbox{$\,$M$_\odot$} }
\def\Lsun{\hbox{$\,$L$_\odot$} }
\def\Teff{\hbox{$\,T_{\rm eff}$} }
\def\alog#1{\times 10^{#1}}
\def\rin{\hbox{$r_{\rm in}$} }
\def\rout{\hbox{$r_{\rm out}$} }

\def\lstar{\ifmmode{\Lambda^*}\else\hbox{$\Lambda^*$}\fi} 
\def\Lstar{\ifmmode{\Lambda^*}\else\hbox{$\Lambda^*$}\fi} 
\def\Rop{\ifmmode{[R_{ij}]}\else\hbox{$[R_{ij}]$}\fi}
\def\Rij{\Rop}
\def\Rji{\ifmmode{[R_{ji}]}\else\hbox{$[R_{ji}]$}\fi}
\def\Rstar{\ifmmode{[R_{ij}^*]}\else\hbox{$[R_{ij}^*]$}\fi}
\def\Rijstar{\Rstar}
\def\Rjistar{\ifmmode{[R_{ji}^*]}\else\hbox{$[R_{ji}^*]$}\fi}
\def\DRji{\ifmmode{[\Delta R_{ji}]}\else\hbox{$[\Delta R_{ji}]$}\fi}
\def\DRij{\ifmmode{[\Delta R_{ij}]}\else\hbox{$[\Delta R_{ij}]$}\fi}

\def\Jb{{\bar J}}
\def\Jnew{{\bar J_{\rm new}}}
\def\Jold{{\bar J_{\rm old}}}
\def\Jfs{{\bar J_{\rm fs}}}
\def\Snew{{S_{\rm new}}}
\def\Sold{{S_{\rm old}}}
\def\Amat{\mat{A}}             

\def\ns{\ifmmode{N_{\rm s}}          
        \else\hbox{$N_{\rm s}$}\fi}
\def\ion#1{\hbox{ #1}}         

\def\peq{\mathbin{\hbox{$+$}\hbox{$=$}}}

\def\mat#1{{\bf #1}}     
\def\vek#1{{#1}}         

\newcount\eqcount
\eqcount=0
\def
  \nummer{
    \global\advance\eqcount by 1
    (\the\eqcount)
  }

\def
  \numadv{
    \global\advance\eqcount by 1
  }

\def
   \numout#1{
     (\the\eqcount #1)
  }

\def\ivek#1#2{\ifmmode{\vek{I}^{#1}_{#2}}
        \else\hbox{$\vek{I}^{#1}_{#2}$}\fi}

\def\ip#1{\ivek{+}{#1}}      
\def\im#1{\ivek{-}{#1}}      

\def\tmat#1#2{\ifmmode{\mat{t}^{#1}_{#2}}
        \else\hbox{$\mat{t}^{#1}_{#2}$}\fi}
\def\rmat#1#2{\ifmmode{\mat{r}^{#1}_{#2}}
        \else\hbox{$\mat{r}^{#1}_{#2}$}\fi}
\def\bvek#1#2{\ifmmode{\beta^{#1}_{#2}}
        \else\hbox{$\beta^{#1}_{#2}$}\fi}

\def\tpi#1{\tmat{+}{#1}}
\def\tmi#1{\tmat{-}{#1}}
\def\rmi#1{\rmat{-}{#1}}
\def\rpi#1{\rmat{+}{#1}}
\def\bpi#1{\bvek{+}{#1}}
\def\bmi#1{\bvek{-}{#1}}

\def\tp{\tmat{+}{}}          
\def\tm{\tmat{-}{}}          
\def\rmm{\rmat{-}{}}         
\def\rp{\rmat{+}{}}          
\def\bp{\bvek{+}{}}          
\def\bm{\bvek{-}{}}          
\def\tpm{\tmat{\pm}{}}       
\def\rpm{\rmat{\pm}{}}       
\def\bpm{\bvek{\pm}{}}       

\def\lp{\ifmmode{\lambda^+_\tau}           
        \else\hbox{$\lambda^+_\tau$}\fi}
\def\lm{\ifmmode\lambda^-_\tau             
        \else\hbox{$\lambda^-_\tau$}\fi}

%
%
%
%



\def\aasref@jnl#1{{\rm #1}}

\def\aj{\aasref@jnl{AJ}}                   
\def\araa{\aasref@jnl{ARA\&A}}             
\def\apj{\aasref@jnl{ApJ}}                 
\def\apjl{\aasref@jnl{ApJ}}                
\def\apjs{\aasref@jnl{ApJS}}               
\def\ao{\aasref@jnl{Appl.~Opt.}}           
\def\apss{\aasref@jnl{Ap\&SS}}             
\def\aap{\aasref@jnl{A\&A}}                
\def\aapr{\aasref@jnl{A\&A~Rev.}}          
\def\aaps{\aasref@jnl{A\&AS}}              
\def\azh{\aasref@jnl{AZh}}                 
\def\baas{\aasref@jnl{BAAS}}               
\def\jrasc{\aasref@jnl{JRASC}}             
\def\memras{\aasref@jnl{MmRAS}}            
\def\mnras{\aasref@jnl{MNRAS}}             
\def\pra{\aasref@jnl{Phys.~Rev.~A}}        
\def\prb{\aasref@jnl{Phys.~Rev.~B}}        
\def\prc{\aasref@jnl{Phys.~Rev.~C}}        
\def\prd{\aasref@jnl{Phys.~Rev.~D}}        
\def\pre{\aasref@jnl{Phys.~Rev.~E}}        
\def\prl{\aasref@jnl{Phys.~Rev.~Lett.}}    
\def\pasp{\aasref@jnl{PASP}}               
\def\pasj{\aasref@jnl{PASJ}}               
\def\qjras{\aasref@jnl{QJRAS}}             
\def\skytel{\aasref@jnl{S\&T}}             
\def\solphys{\aasref@jnl{Sol.~Phys.}}      
\def\sovast{\aasref@jnl{Soviet~Ast.}}      
\def\ssr{\aasref@jnl{Space~Sci.~Rev.}}     
\def\zap{\aasref@jnl{ZAp}}                 
\def\nat{\aasref@jnl{Nature}}              
\def\iaucirc{\aasref@jnl{IAU~Circ.}}       
\def\aplett{\aasref@jnl{Astrophys.~Lett.}} 
\def\apspr{\aasref@jnl{Astrophys.~Space~Phys.~Res.}}
\def\bain{\aasref@jnl{Bull.~Astron.~Inst.~Netherlands}} 
\def\fcp{\aasref@jnl{Fund.~Cosmic~Phys.}}  
\def\gca{\aasref@jnl{Geochim.~Cosmochim.~Acta}}   
\def\grl{\aasref@jnl{Geophys.~Res.~Lett.}} 
\def\jcp{\aasref@jnl{J.~Chem.~Phys.}}      
\def\jgr{\aasref@jnl{J.~Geophys.~Res.}}    
\def\jqsrt{\aasref@jnl{J.~Quant.~Spec.~Radiat.~Transf.}}
\def\memsai{\aasref@jnl{Mem.~Soc.~Astron.~Italiana}}
\def\nphysa{\aasref@jnl{Nucl.~Phys.~A}}   
\def\physrep{\aasref@jnl{Phys.~Rep.}}   
\def\physscr{\aasref@jnl{Phys.~Scr}}   
\def\planss{\aasref@jnl{Planet.~Space~Sci.}}   
\def\procspie{\aasref@jnl{Proc.~SPIE}}   

\let\astap=\aap
\let\apjlett=\apjl
\let\apjsupp=\apjs
\let\applopt=\ao

\baselineskip=12pt

\title{A 3D radiative transfer framework: III. periodic boundary conditions}

\titlerunning{3D radiative transfer framework III}
\authorrunning{Hauschildt and Baron}
\author{Peter H. Hauschildt\inst{1} and E.~Baron\inst{1,2,3}}

\institute{
Hamburger Sternwarte, Gojenbergsweg 112, 21029 Hamburg, Germany;
yeti@hs.uni-hamburg.de 
\and
Dept. of Physics and Astronomy, University of
Oklahoma, 440 W.  Brooks, Rm 100, Norman, OK 73019 USA;
baron@ou.edu
\and
Computational Research Division, Lawrence Berkeley National Laboratory, MS
50F-1650, 1 Cyclotron Rd, Berkeley, CA 94720-8139 USA
}

\date{Received date \ Accepted date}

\abstract
{}
{We present a general method to solve radiative transfer problems including
scattering in the continuum as well as in lines in 3D configurations with 
periodic boundary conditions.
}
{The scattering problem for line transfer is solved via means of an
  operator splitting (OS) technique. The formal solution
  is based on a full characteristics method. The approximate
  $\Lambda$ operator is constructed considering nearest
  neighbors exactly. The code is parallelized over both wavelength and solid angle
  using the MPI library.
}
{We present the results of several test cases with different values of
  the thermalization parameter and two choices for the temperature
  structure. The results are directly compared to 1D plane parallel
  tests. The 3D results agree very
  well with the well-tested 1D calculations.
}
{Advances in modern computers will make realistic 3D radiative transfer
  calculations possible in the near future. Our current code scales to
very large numbers of processors, but requires larger memory per
processor at high spatial resolution.}


\maketitle

\section{Introduction}

Interest in 3-D radiative transfer in stellar atmospheres has grown
with the calculations of 
Asplund and collaborators
\citep{ANTS99,ANTS00,AspIII00,AGS02,GAS07}. This work has indicated
that the solar oxygen abundance needs to be revised downward. However,
the revised abundances are difficult to reconcile with
helioseismological results \citep[see][and references therein]{BAPR08}.
The work of Asplund et al.\ is based on comparisons of synthetic
spectra produced by formal solutions of hydrodynamical models of solar
convection. We present a framework for solving the full scattering
problem that is applicable to hydrodynamical calculations of stellar
atmospheres.
\citet[][hereafter: Paper I]{hb06} and
\citet[][hereafter: Paper II]{bh07}  described a framework for
the solution of the radiative transfer equation for scattering continua
and lines
in 3D (when we say 3D we mean three spatial dimensions, plus three
momentum dimensions) for 
the time independent, static case. In the 3rd paper of this series
we apply these methods to problems with period boundary conditions
which typically arise in radiation-hydrodynamical simulations
of convective atmospheres. In such calculations the radiation
transport has to be simplified compared to the full problem in 
order to keep the calculations tractable. However, a full solution of
the scattering line problem is needed for comparison and post-processing
of the structures.

{We describe our method, its rate of convergence, and present
  comparisons to our 
  well-tested 1-D calculations.}

\section{Method}

In the following discussion we use  notation of Papers I and II.  The
basic framework and the methods used for the formal solution and the solution
of the scattering problem via operator splitting are discussed in detail in
Papers I and II and will thus not be repeated here. 

In the following we assume (without restriction) that we have periodic
boundary conditions in the $x$ and $y$ coordinates, and for the $z$
coordinate that the `bottom' (large optical depth) is at $z=z_{\rm
  min}$ and the `top' (interface to empty space) is at $z=z_{\rm
  max}$. The implementation of the periodic boundary conditions within
our framework is simple:
We use a ``full characteristics'' approach that { completely} tracks a set
of characteristics of the radiative transfer equation { from the outer
boundary} through the computational domain { to their exit voxel and takes
care that each voxel is hit by at least one characteristic per solid angle}.
One characteristic is
started on each boundary voxel (in this case these are the planes $z=
z_{\rm min}$ and $z=z_{\rm max}$) and then tracked until it leaves the
other boundary.  The direction of a bundle of full characteristics is
determined by a set of solid angles $(\theta,\varphi)$ which
correspond to a normalized momentum space vector $(p_x,p_y,p_z)$.  The
periodic boundary conditions are simply implemented as a wrap-around
(e.g., passing $x_{\rm max}$ for $p_x>0$ wraps around to $x_{\min}$)
and continuing of the characteristic until it leaves at the $z$
boundary. Characteristics with very small $|p_z|$ would require a
large number of wrap-arounds { (and eventually would lead
to infinitely long characteristics)}, therefore, we limit the number of
wrap-arounds per voxel to a prescribed value, typically around 16 (tests
have shown that larger values do not affect the results, values as
small as 4 are usable in plane-parallel tests).  The code is
parallelized as described in Paper II.

\section{Plane-parallel tests}

\subsection{Testing environment}
We use the framework discussed in Paper I and II as the baseline for the
line transfer problems discussed in this paper. 
Our basic setup is similar to that discussed in Paper II.  
Periodic boundary conditions (PBCs) are realized in a plane parallel 
slab. We use PBCs on the $x$ and $y$ axes, $z_{\rm max}$ is at the 
outside boundary, $z_{\rm min}$ the inside boundary. The slab has
a finite optical depth in the $z$ axis.
The grey continuum opacity is parameterized by a power law in the continuum optical
depth $\tstd$ in the $z$ axis. The basic model parameters are
\begin{enumerate}
\item thickness of the slab,  $z_{\rm max}- z_{\rm min} = 10^7\,$cm
\item Minimum optical depth in the continuum, $\t_{\rm std}^{\rm min} =
10^{-8}$ and maximum optical depth in the continuum, $\t_{\rm std}^{\rm
max} = 10$.
\item Constant temperatures (in all axes), $T=10^4$~K
\item Outer boundary condition, $I_{\rm bc}^{-} \equiv 0$ and diffusion
inner boundary condition for all wavelengths.
\item Parameterized coherent \& isotropic continuum scattering by
defining
\[
\chi_c = \epsilon_c \kappa_c + (1-\epsilon_c) \sigma_c
\]
with $0\le \epsilon_c \le 1$. 
$\kappa_c$ and $\sigma_c$ are the
continuum absorption and scattering coefficients.
\end{enumerate}

The line of the simple 2-level model atom is parameterized by the ratio of the
profile averaged  line opacity $\chi_l$ to the continuum opacity $\chi_c$ and
the line thermalization parameter $\epsilon_l$. For the test cases presented
below, we have used $\epsilon_c=1$ and a constant temperature and thus a
constant thermal part of the source function for simplicity (and to save
computing time) and set $\chi_l/\chi_c = 10^6$ to simulate a strong line, with
varying $\epsilon_l$ (see below). With this setup, the optical depths as seen
in the line range from $10^{-2}$ to $10^6$. We use 32 wavelength points to model
the full line profile, including wavelengths outside the line for the
continuum.  We did not require the line to thermalize at the center of the test
configurations, this is a typical situation one encounters in a full 3D
configurations as the location (or even existence) of the
thermalization depths becomes more 
ambiguous than in the 1D case.

The slab is mapped onto a Cartesian grid.
For the test calculations we use voxel grids with the same
number of spatial points in each direction (see below). The
solid angle space was discretized in $(\theta,\phi)$ with 
$n_\theta=n_\phi$ if not stated otherwise. In the following 
we discuss the results of various tests. In all tests we use
the full characteristic method for the 3D RT solution as described above.
Unless otherwise stated,
the tests were run on parallel computers using 128 CPUs. For the 3D solver
we use $n_x=n_y=n_z=2*32+1=65$ points along each axis.
The solid angle space discretization uses $n_\theta=n_\phi=64$ points.

\subsection{Results}

\begin{figure}
\centering
\resizebox{\hsize}{!}{\includegraphics[angle=90]{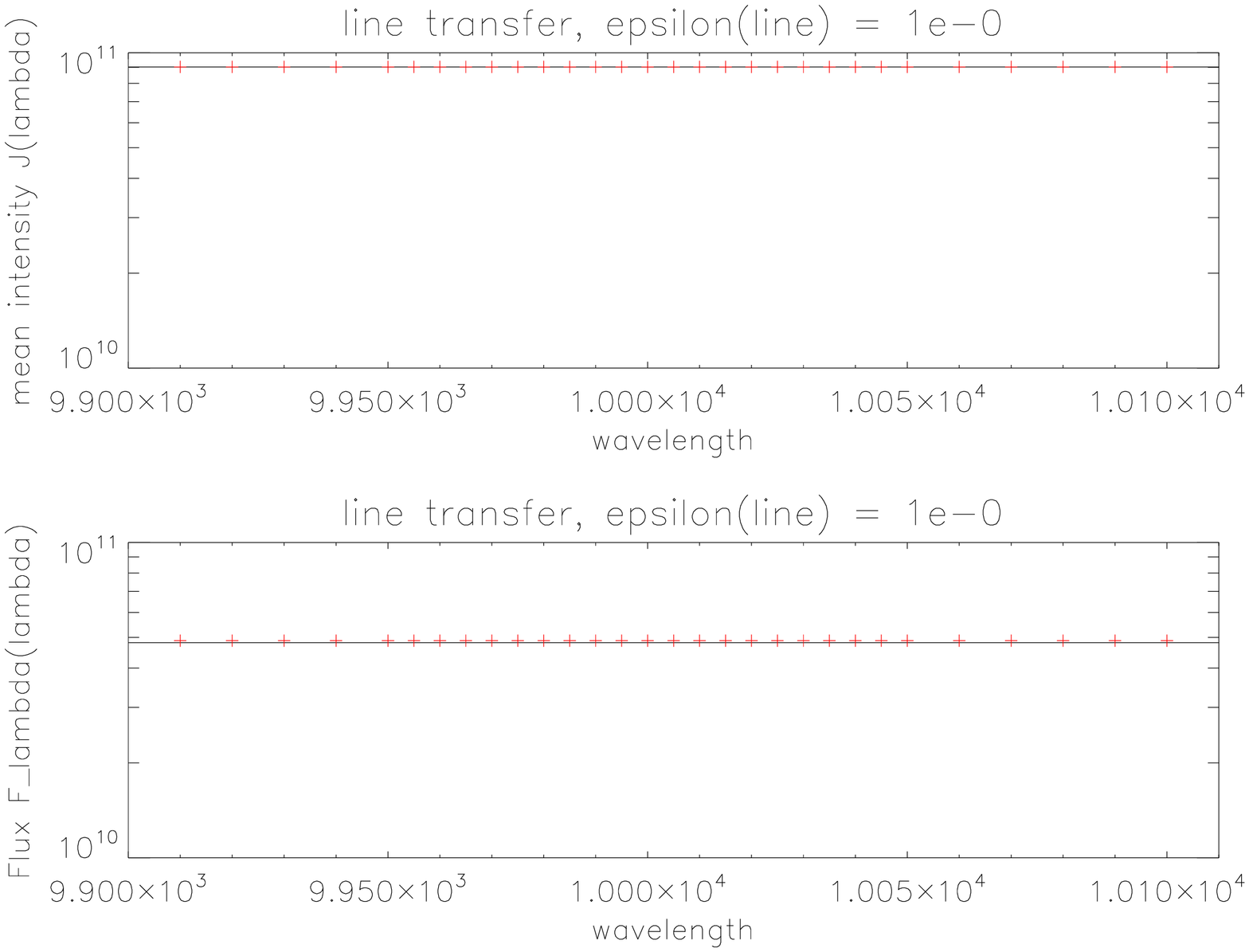}}
\caption{\label{fig:LTE:spec} The mean intensity $J$ and the $z$ component
of the radiation flux $F$ at $\tstd=0$ as function of wavelength. The 
$+$ symbols are the comparison results with the 1D solver, the full lines
the results from the 3D PBC solution. The results are for $\epsilon_l=1$ 
and constant temperatures.}
\end{figure}

\begin{figure}
\centering
\resizebox{\hsize}{!}{\includegraphics[angle=90]{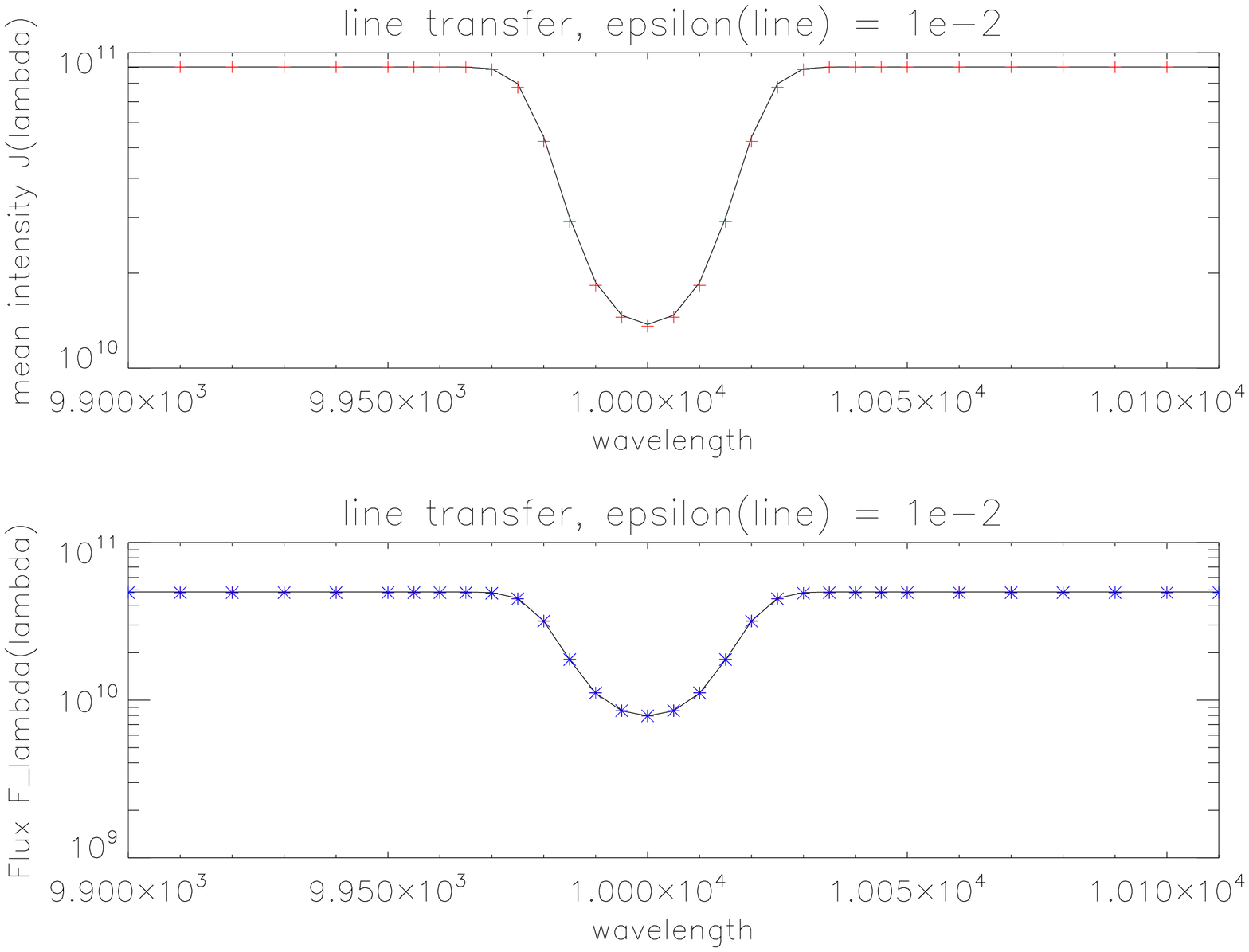}}
\caption{\label{fig:2:spec} The mean intensity $J$ and the $z$ component
of the radiation flux $F$ at $\tstd=0$ as function of wavelength. The 
$+$ symbols are the comparison results with the 1D solver, the full lines
the results from the 3D PBC solution. The results are for $\epsilon_l=10^{-2}$ 
and constant temperatures.}
\end{figure}

\begin{figure}
\centering
\resizebox{\hsize}{!}{\includegraphics[angle=90]{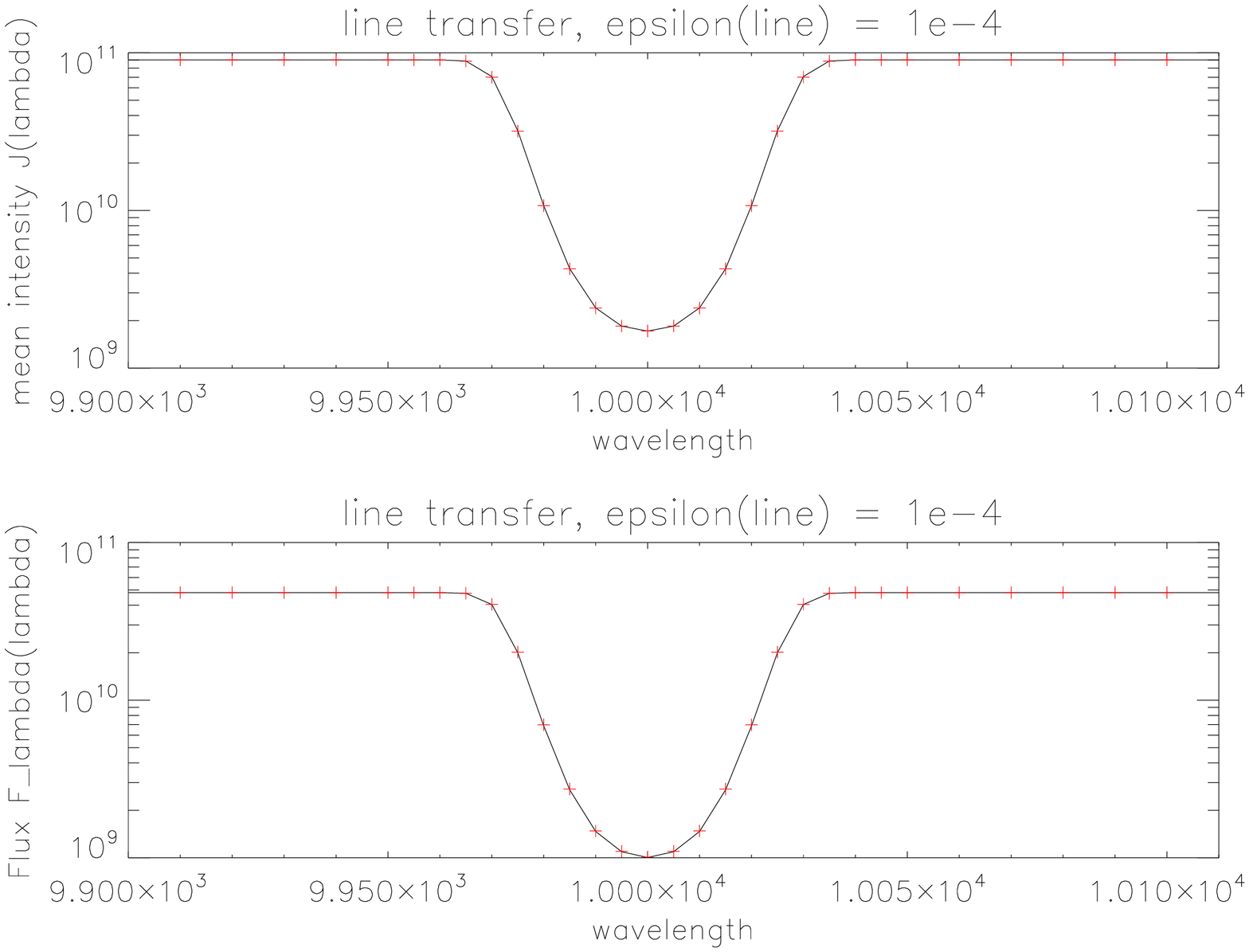}}
\caption{\label{fig:4:spec} The mean intensity $J$ and the $z$ component
of the radiation flux $F$ at $\tstd=0$ as function of wavelength. The 
$+$ symbols are the comparison results with the 1D solver, the full lines
the results from the 3D PBC solution. The results are for $\epsilon_l=10^{-4}$ 
and constant temperatures.}
\end{figure}

\begin{figure}
\centering
\resizebox{\hsize}{!}{\includegraphics[angle=90]{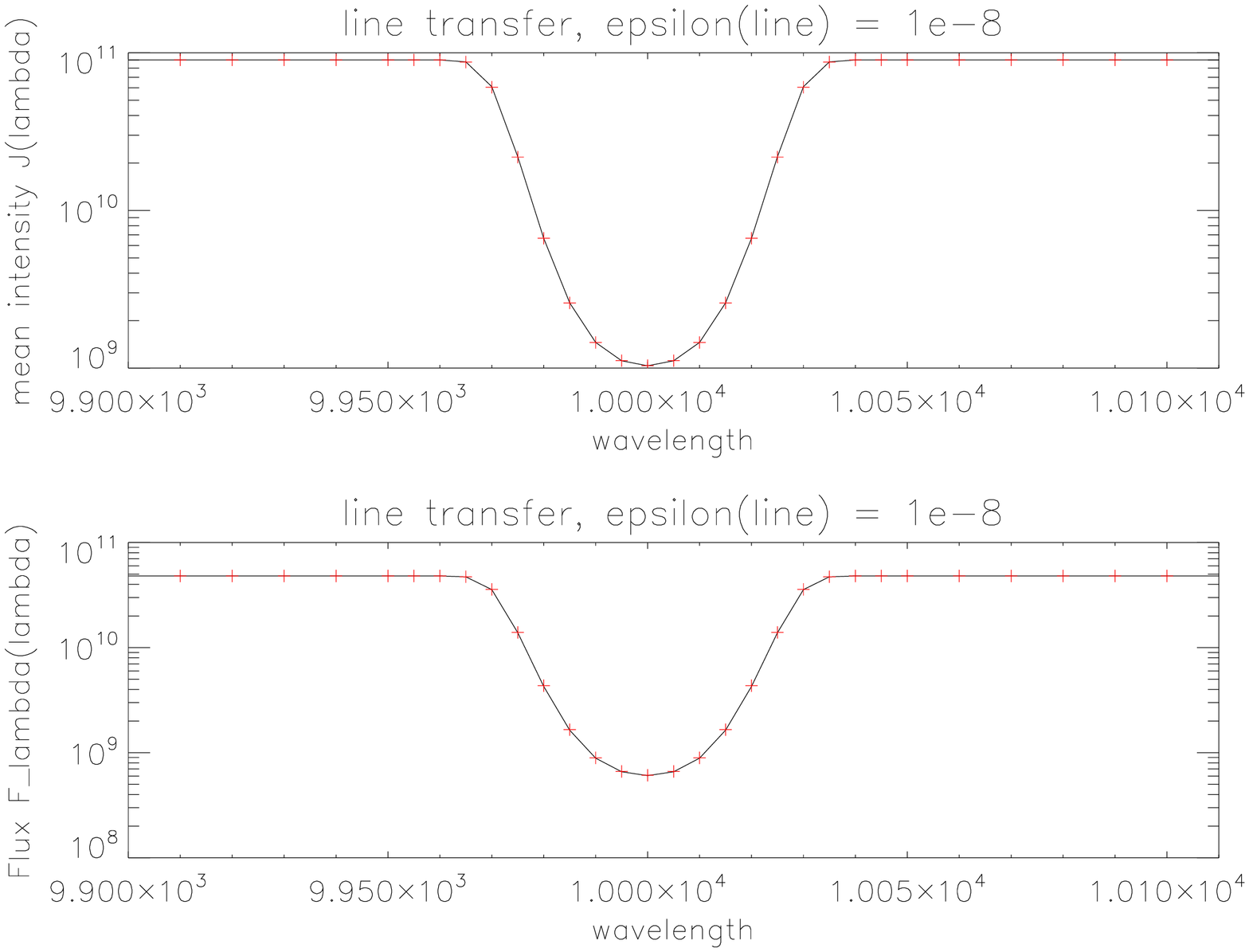}}
\caption{\label{fig:8:spec} The mean intensity $J$ and the $z$ component
of the radiation flux $F$ at $\tstd=0$ as function of wavelength. The 
$+$ symbols are the comparison results with the 1D solver, the full lines
the results from the 3D PBC solution. The results are for $\epsilon_l=10^{-8}$ 
and constant temperatures.}
\end{figure}

We test the accuracy of the 3D PBC solution by comparing it to
the results of the 1D code for several line scattering parameters.
The 1D solver uses 64 depth points, distributed logarithmically in
optical depth.  Figures~\ref{fig:LTE:spec}--\ref{fig:8:spec} 
show the mean intensities $\Jb$ at $\tstd=0$ and the $z$ component of
the emergent flux $F$ as function of wavelength for both the 1D ($+$
symbols) and the 3D solver. The agreement is excellent for all values
of $\epsilon_l$ from unity to $10^{-8}$, indicating that the 3D code
produces an accurate solution even for extreme cases of
line scattering.
{ In the case with $\epsilon_l=10^{-8}$ the continuum
processes lead to earler thermalization than the classical 
approximation $J\propto \epsilon^{1/2}$ as the line strength is 
limited compared to the continuum. This behavior is the same as in the
1D plane-parallel comparison case.}
The convergence rate of the line source function { (here used
together with Ng acceleration)} is the
same as discussed in Paper II, in the case of $\epsilon_l=10^{-8}$ the
3D code needed 29 iterations with the nearest-neighbor $\lstar$ to
reach a relative accuracy of $10^{-8}$ using the simple starting guess
$S=B$.
{ The nearest-neighbor $\lstar$ does allow stopping the 
iterations earlier than a diagonal (local) $\lstar$ due to the 
improved convergence rate (see paper I). This can easily cut
the number of iterations by factors of two or more, even greater
savings are possible if the accuracy limit is relaxed.}

In addition to the mean intensities, we checked that the flux 
vectors $\vec F$ have vanishing components in the $x$ and $y$ 
directions, typically $\max(|F_x|,|F_y|)/|F_z| \le 10^{-13}$ in
all voxels. We stress that this result is the result of the
calculations and is not forced by the numerical scheme.

\section{Tests with 3D structures}

For a test with a computed 3D structure, we obtained an example snapshot
structure from H-G.\ Ludwig \cite[]{2007A&A...473L...9C,2004A&A...414.1121W} of a 
radiation-hydrodynamical simulation of convection in the solar atmosphere. The
radiation transport calculations were performed with a total of $141\times 141
\times 151$ grid points in $x$,$y$, and $z$, respectively, for a total of
$3\,002\,031$ voxels, the periodic boundary conditions were set in the
(horizontal) $x,y$ plane. The transport equation is solved
for $n_\theta=16$ and $n_\phi=32$ solid angle points, so that a total
of about $1.5\alog{9}$ intensities are calculated for each iteration
and wavelength point. For the tests described here, we are only using the temperature--density
structure of the hydro model and ignore the velocity field for the simple tests
presented here. We set the continuum opacity proportional to the density $\rho$
by choosing a rough temperature independent estimate for the Rosseland mean
opacity per unit mass of $0.1\,$cm$^2$/g and parameterizing the line in the
same form as discussed above and in Paper II with a line total (wavelength
integrated) opacity  of $100$ times the local continuum and 32 wavelength
points distributed over the full line profile.

\subsection{Results}

\begin{figure}
\centering
\resizebox{\hsize}{!}{\includegraphics[angle=90]{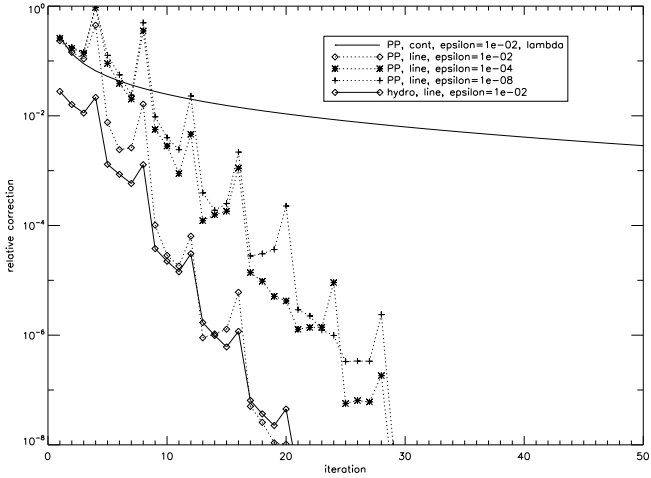}}
\caption{\label{fig:convergence} Convergence rates of the 3D transfer for
line transfer with plane-parallel test structures (label 'PP') and the 3D
hydro structure (label 'hydro'). For comparison, the convergence of
the $\Lambda$ iteration for plane-parallel continuum transfer is 
also shown.}
\end{figure}

We ran a number of line transfer tests with $\epsilon_l=1$, $10^{-2}$
and $10^{-4}$.  The convergence rates for the two scattering cases are
shown in Fig.~\ref{fig:convergence} together with the convergence
rates for a small plane parallel test model and the results of the
$\Lambda$ iteration for continuum transfer (for computer time 
reasons) for a continuum $\epsilon_c=10^{-2}$. The convergence rate
for the plane-parallel tests and the hydro model are remarkably
similar, the $\Lstar$ operator delivers very reasonable and
practically usable convergence rates.

\subsubsection{Images}

Figures~\ref{fig:vis:1} and \ref{fig:vis:2} show visualizations
of the results for 3D continuum transfer. The RT problem was solved for 
$\epsilon=1$ (left panels) and $10^{-2}$ (right panels) and a formal
solution with the converged source functions was computed for given 
viewing angles.  The graphs are actual images
of the intensities as they would be seen by an external observer 
different angles. The visible surface is to the left,
the 'sides' of the computational box could not be seen by an observer and
are shown for information only. The effect of scattering on
the images is  similar to terrestrial fog in that it
reduces the contrast of visible features; even moderate scattering
of $\epsilon_c=10^{-2}$ significantly reduces visibility. The limb darkening
is also clearly visible in the figures. 

{
Figures~\ref{fig:line:vis:1} to \ref{fig:line:vis:4} show images generated for
the results of the line transfer solution.  Three panels show results for
individual wavelength (continuum, line wing and line center) and a composite
image. The images are significantly different for these wavelengths. The line
scattering produces a similar 'fog effect' as the scattering in the continuum
transfer model, however, the images appear not that different. While one might
expect that the line images would look vastly different from the continuum
visualization, part of the similarity is due to  the fact that they were scaled
individually in order to highlight the differences in structure between the
wavelengths rather than comparing them on a absolute scale. The composite image
(best viewed in color available in the online version of this paper) shows the
differences in the visible structures between the different wavelengths.
}

\subsubsection{Limb darkening and contrast}

\begin{figure*}
\centering
\begin{minipage}{0.45\hsize}
\includegraphics[width=0.8\hsize,angle=90]{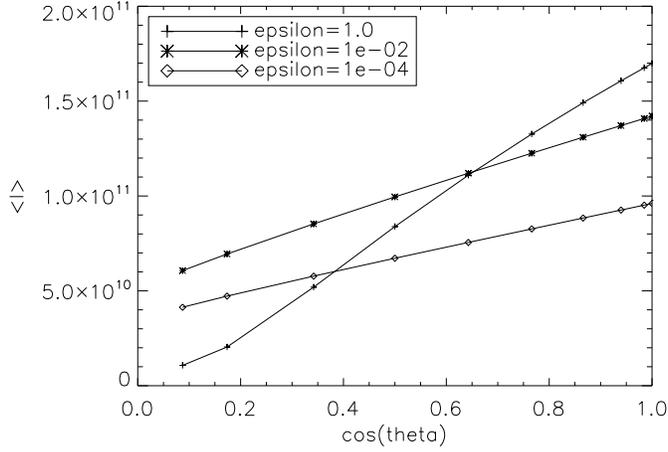}
\end{minipage}
\begin{minipage}{0.45\hsize}
\includegraphics[width=0.8\hsize,angle=90]{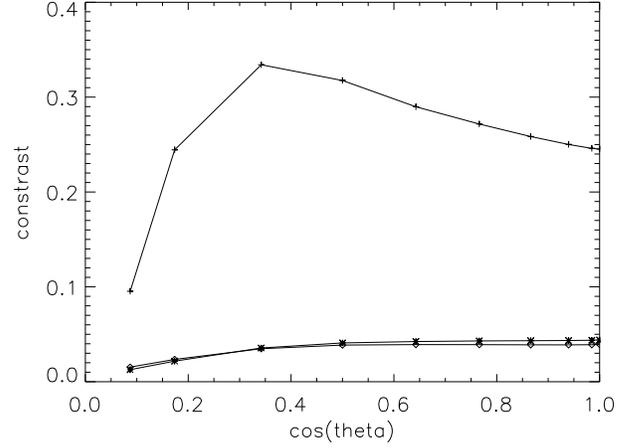}
\end{minipage}
\caption{\label{fig:contrast} Limb darkening (left panel) and
contrast $\sqrt{\langle(I-\langle I\rangle )^2\rangle )}/\langle I\rangle $ for 3D continuum 
transfer with the hydro structure.}
\end{figure*}

In Fig.~\ref{fig:contrast} we show the limb darkening and 
contrast for a continuum test case with different
values of $\epsilon$. To compute the limb darkening, we
calculate the intensity average $<I>$ over the visible surface
for different values of $\cos(\theta)$ where $\theta$ is the 
angle between the observer and the normal to the surface. We
similarly calculate the contrast as $\sqrt{<(I-<I>)^2>)}/<I>$
over the visible surface for different $\theta$. The absolute
values of the limb darkening and the contrast depend strongly 
on $\epsilon$, scattering dramatically reduces the 
contrast and 'flattens' the limb darkening law. Overall, the limb
darkening is nearly linear in $\cos(\theta)$, as would be
expected from a plane parallel atmosphere with grey temperature
structure. 

\section{Conclusions}

Using rather difficult plane parallel test problems, we have shown
that our 3D full-characteristics method gives very good results when
compared to our well-tested 1D code. The periodic boundary conditions
method discussed here is particularly well-suited to 3-D
hydrodynamical simulations of convection in stellar atmospheres and in
future work we will compare our results to observations as well as to
previous calculations.  The results for the computed 3D structure show
that $\Lstar$ leads also to good convergence for a true 3D structure,
with convergence rates that are comparable to the simple test cases
(see also Papers I and II).

\begin{acknowledgements}
  This work was supported in part by by NASA grants NAG5-3505 and
  NAG5-12127,  NSF grants AST-0307323, and
  AST-0707704, and US DOE Grant DE-FG02-07ER41517, as well as 
  DFG GrK 1351 and SFB 676.
  Some of the
  calculations presented here were performed at the H\"ochstleistungs
  Rechenzentrum Nord (HLRN); at the NASA's Advanced Supercomputing
  Division's Project Columbia, at the Hamburger Sternwarte Apple G5
  and Delta Opteron clusters financially supported by the DFG and the
  State of Hamburg; and at the National Energy Research Supercomputer
  Center (NERSC), which is supported by the Office of Science of the
  U.S.  Department of Energy under Contract No. DE-AC03-76SF00098.  We
  thank all these institutions for a generous allocation of computer
  time.
\end{acknowledgements}

\bibliography{refs,baron,stars,rte}

\begin{thebibliography}{10}
\expandafter\ifx\csname natexlab\endcsname\relax\def\natexlab#1{#1}\fi

\bibitem[{{Asplund}(2000)}]{AspIII00}
{Asplund}, M. 2000, \aap, 359, 755

\bibitem[{{Asplund} {et~al.}(2005){Asplund}, {Grevesse}, \& {Sauval}}]{AGS02}
{Asplund}, M., {Grevesse}, N., \& {Sauval}, A.~J. 2005, in ASP Conf. Ser. 336:
  Cosmic Abundances as Records of Stellar Evolution and Nucleosynthesis, ed.
  T.~G. {Barnes}, III \& F.~N. {Bash}, 25

\bibitem[{{Asplund} {et~al.}(1999){Asplund}, {Nordlund}, {Trampedach}, \&
  {Stein}}]{ANTS99}
{Asplund}, M., {Nordlund}, {\AA}., {Trampedach}, R., \& {Stein}, R.~F. 1999,
  \aap, 346, L17

\bibitem[{{Asplund} {et~al.}(2000){Asplund}, {Nordlund}, {Trampedach}, \&
  {Stein}}]{ANTS00}
{Asplund}, M., {Nordlund}, {\AA}., {Trampedach}, R., \& {Stein}, R.~F. 2000,
  \aap, 359, 743

\bibitem[{Baron \& Hauschildt(2007)}]{bh07}
Baron, E. \& Hauschildt, P.~H. 2007, A\&A, 468, 255

\bibitem[{{Basu} \& {Antia}(2008)}]{BAPR08}
{Basu}, S. \& {Antia}, H.~M. 2008, \physrep, 457, 217

\bibitem[{{Caffau} {et~al.}(2007){Caffau}, {Steffen}, {Sbordone}, {Ludwig}, \&
  {Bonifacio}}]{2007A&A...473L...9C}
{Caffau}, E., {Steffen}, M., {Sbordone}, L., {Ludwig}, H.-G., \& {Bonifacio},
  P. 2007, \aap, 473, L9

\bibitem[{{Grevesse} {et~al.}(2007){Grevesse}, {Asplund}, \& {Sauval}}]{GAS07}
{Grevesse}, N., {Asplund}, M., \& {Sauval}, A.~J. 2007, Space Science Reviews,
  130, 105

\bibitem[{Hauschildt \& Baron(2006)}]{hb06}
Hauschildt, P.~H. \& Baron, E. 2006, A\&A, 451, 273

\bibitem[{{Wedemeyer} {et~al.}(2004){Wedemeyer}, {Freytag}, {Steffen},
  {Ludwig}, \& {Holweger}}]{2004A&A...414.1121W}
{Wedemeyer}, S., {Freytag}, B., {Steffen}, M., {Ludwig}, H.-G., \& {Holweger},
  H. 2004, \aap, 414, 1121

\end{thebibliography}

\clearpage

\begin{figure}
\centering
\caption{\label{fig:vis:1} Visualization of the results for continuum
3D radiation transfer for $\epsilon_c=1$ (left panel) and $10^{-2}$ (right panel).
The images are intensities in the directions $\phi=25\deg$ and $\theta=0\deg$
(top row) and $\theta=40\deg$ (bottom row). An observer would only see the 
left face of the cube (inside the indicated area), the other sides of the cube 
are shown for clarity and are actually invisible due to the periodic boundary
conditions. The scaling of the intensities is the same within each
column but different for the left and right columns.}
\end{figure}

\begin{figure}
\centering
\caption{\label{fig:vis:2} Same as Fig.~\ref{fig:vis:1}, but for 
$\theta=60\deg$ (top row) and $\theta=80\deg$ (bottom row). The 
scaling of the intensities is as in Fig.~\ref{fig:vis:1}. The effect of limb
darkening is clearly visible in this figure.}
\end{figure}

\begin{figure}
\centering
\caption{\label{fig:line:vis:1} Visualization of the results for the line
3D radiation transfer with $\epsilon_l=1$.
The images are intensities in the directions $\phi=25\deg$ and $\theta=0\deg$.
The top left panel is the image in the continuum, the top right panel the image 
at the line center, the bottom left panel the image in the line wing, the bottom
right panel is a composite image.}
\end{figure}

\begin{figure}
\centering
\caption{\label{fig:line:vis:2} Visualization of the results for the line
3D radiation transfer with $\epsilon_l=10^{-4}$.
The images are intensities in the directions $\phi=25\deg$ and $\theta=0\deg$.
The top left panel is the image in the continuum, the top right panel the image 
at the line center, the bottom left panel the image in the line wing, the bottom
right panel is a composite image.}
\end{figure}

\begin{figure}
\centering
\caption{\label{fig:line:vis:3} Visualization of the results for the line
3D radiation transfer with $\epsilon_l=1$.
The images are intensities in the directions $\phi=25\deg$ and $\theta=50\deg$.
The top left panel is the image in the continuum, the top right panel the image 
at the line center, the bottom left panel the image in the line wing, the bottom
right panel is a composite image.}
\end{figure}

\begin{figure}
\caption{\label{fig:line:vis:4} Visualization of the results for the line
3D radiation transfer with $\epsilon_l=10^{-4}$.
The images are intensities in the directions $\phi=25\deg$ and $\theta=50\deg$.
The top left panel is the image in the continuum, the top right panel the image 
at the line center, the bottom left panel the image in the line wing, the bottom
right panel is a composite image.}
\end{figure}

\end{document}